\documentclass[conference, letterpaper]{IEEEtran}
\IEEEoverridecommandlockouts
\usepackage{cite}
\usepackage{amsmath,amssymb,amsfonts}
\usepackage{algorithmic}
\usepackage{graphicx}
\usepackage{textcomp}
\usepackage{xcolor}

\usepackage[utf8]{inputenc} 
\usepackage{dsfont} 
\usepackage{caption}
\usepackage{subcaption}
\usepackage{url}

\usepackage{etoolbox} 

\newcommand{\scalar}[2]{\left\langle#1,#2\right\rangle}
\newcommand{\Zqn}[0]{\mathbb{Z}_q^n}
\newcommand{\norm}[1]{\left\lVert#1\right\rVert}

\newtheorem{defn}{Definition}

\def\BibTeX{{\rm B\kern-.05em{\sc i\kern-.025em b}\kern-.08em
    T\kern-.1667em\lower.7ex\hbox{E}\kern-.125emX}}
\begin{document}

\newtoggle{full}
\toggletrue{full} 

\title{On the Sample Complexity of solving LWE using BKW-Style Algorithms
}

\author{\IEEEauthorblockN{Qian Guo, Erik Mårtensson and Paul Stankovski Wagner}
\IEEEauthorblockA{\textit{Department of Electrical and Information Technology} \\
Lund University, Lund, Sweden \\
Email: \{qian.guo, erik.martensson, paul.stankovski\_wagner\}@eit.lth.se}
}

\maketitle

\begin{abstract}
	The Learning with Errors (LWE) problem receives much attention in cryptography, mainly due to its fundamental significance in post-quantum cryptography. Among its solving algorithms, the Blum-Kalai-Wasserman (BKW) algorithm, originally proposed for solving the Learning Parity with Noise (LPN) problem, performs well, especially for certain parameter settings with cryptographic importance. The BKW algorithm consists of two phases, the reduction phase and the solving phase.
	
	In this work, we study the performance of distinguishers used in the solving phase. We show that the Fast Fourier Transform (FFT) distinguisher from Eurocrypt'15 has the same sample complexity as the optimal distinguisher, when making the same number of hypotheses. We also show that it performs much better than theory predicts and introduce an improvement of it called the pruned FFT distinguisher. Finally, we indicate, via extensive experiments, that the sample dependency due to both LF2 and sample amplification is limited.
\end{abstract}

\iftoggle{full}{%
	
}{
	\textit{A full version of this paper is accessible at~\cite{gms2021}.}
}

 

\section{Introduction}
\label{sec:intro}

Post-quantum cryptography  studies replacements of cryptographic primitives based on the factoring or discrete-log problem, since both can be efficiently solved by a quantum computer~\cite{FOCS:Shor94}. Lattice-based cryptography is its main area. In the NIST Post-Quantum Cryptography Standardization~\cite{NISTPQCweb}, 5 out of 7 finalists and 2 out of 8 alternates are lattice-based.
  
The {\em Learning with Errors} (LWE) problem, introduced by Regev~\cite{STOC:Regev05}, is the major problem in lattice-based cryptography. Its average-case hardness can be based on the worst-case hardness of some standard lattice problems, which is extremely interesting in theoretical crypto. The most famous, of its many cryptographic applications, is the design of Fully Homomorphic Encryption (FHE) schemes.
Its binary counterpart, the {\em Learning Parity with Noise} problem (LPN), also plays an significant role in cryptography (see~\cite{C:BFKL93}), especially in light-weight cryptography for very constrained environments such as RFID tags and low-power devices. 

The algorithms for solving LWE can be divided into lattice-based, algebraic, and combinatorial methods. The last class of algorithms all inherit from the famous Blum-Kalai-Wasserman (BKW) algorithm~\cite{STOC:BluKalWas00,DBLP:journals/jacm/BlumKW03}, and are the most relevant to our study. We refer interested readers to~\cite{ConcreteLWE} for concrete complexity estimation for solving LWE instances, and to~\cite{DBLP:journals/dcc/HeroldKM18,DBLP:journals/tit/GuoJMW19} for asymptotic complexity estimations.

The BKW-type algorithms include two phases, the reduction phase and the solving phase. The prior consists of a series of operations, called BKW steps, iteratively reducing the dimension of the problem at the cost of increasing its noise level. At the end of the reduction phase, the original LWE problem is transformed to a new problem with a much smaller dimension. The new problem can be solved efficiently by a procedure called distinguishing in the solving phase.

One of the main challenges in understanding the precise performance of BKW variants on solving the LWE problem comes from the lack of extensive experimental studies, especially on the various distinguishers proposed for the solving phase. Firstly, we have borrowed many heuristics from BKW variants on the LPN problem, but only very roughly or not at all verified them for the LWE problem.
Secondly, the tightness of the nice theoretical bound in~\cite{EC:DucTraVau15} on the sample complexity of the FFT distinguisher also needs to be experimentally checked.
Lastly, a performance comparison of the different known distinguishers is still lacking.


\subsection{Related Work}
\label{sec:Related}

The BKW algorithm proposed by Blum et al.~\cite{STOC:BluKalWas00,DBLP:journals/jacm/BlumKW03} is the first sub-exponential algorithm for solving the LPN problem. Its initial distinguisher, an exhaustive search method in the binary field, recovers one bit of the secret by employing majority voting.  Later, Levieil and Fouque~\cite{SCN:LevFou06} applied the fast Walsh-Hadamard transform (FWHT) technique to accelerate the distinguishing process and recovered a number of secret bits in one pass. They also proposed some heuristic versions and tested these assumptions by experiments. 
In~\cite{EPRINT:Kirchner11} Kirchner proposed a secret-noise transform technique to change the secret distribution to be sparse. This technique is an application of the transform technique proposed in~\cite{C:ACPS09} for solving LWE.
Bernstein and Lange~\cite{EPRINT:BerLan12d} further instantiated an attack on the Ring-LPN problem, a variant of LPN with algebraic ring structures.  In~\cite{AC:GuoJohLon14,DBLP:journals/joc/GuoJL20}, Guo, Johansson, and L\"{o}ndahl proposed a new distinguishing method called subspace hypothesis testing. Though this distinguisher can handle an instance with larger dimension by using covering codes, its inherent nature is still an FWHT distinguisher. 
Improvements of the BKW algorithm were further studied by Zhang et al.~\cite{EC:ZhaJiaWan16} and Bogos-Vaudenay~\cite{AC:BogVau16}. An elaborate survey with experimental results on the BKW algorithm for solving LPN can be found in~\cite{DBLP:journals/ccds/BogosTV16}.

BKW for solving LWE follows a similar research line. Albrecht et al. initiated the study in~\cite{Albrecht2015}. In PKC 2014~\cite{PKC:AFFP14}, a new reduction technique called lazy modulus switching was proposed. In both works, the solving phase uses an exhaustive search approach. In~\cite{EC:DucTraVau15} Duc et al. introduced the fast Fourier transform (FFT) technique in the distinguishing process and bounded the sample complexity theoretically from the Hoeffding inequality. Note that the actual performance regarding the bound is not experimentally verified and the information loss in the FFT distinguisher is unclear. 
There are new reduction methods in~\cite{C:GuoJohSta15,C:KirFou15,AC:GJMS17}, and in~\cite{C:GuoJohSta15}, the authors also proposed a new  method with polynomial reconstruction in the solving phase. This method has the same sample complexity as that of the exhaustive search approach but requires \((q+1)\) FFT operations rather than only one FFT in~\cite{EC:DucTraVau15}. The BKW variants with memory constraints were recently studied in~\cite{C:EssKubMay17,C:EHKMS18,IMA:DelEssMay19}.

        
\subsection{Contributions}  
\label{sec:Contributions}

%
%
%

In the paper, we compare the performances of the known distinguishers empirically. We investigate the performance of the optimal distinguisher and the FFT distinguisher. We also test the sample dependency when using LF2 or sample amplification. We have the following contributions.

\begin{enumerate}
	\item We show that the FFT distinguisher and the optimal distinguisher have the same sample complexity, if we make sure that the distinguishers make the same number of hypotheses. Thus, except for very sparse secrets, the FFT distinguisher is always preferable. This also makes the polynomial reconstruction method of~\cite{C:GuoJohSta15} obsolete.
	\item We indicate that the formula from \cite{EC:DucTraVau15} for the number of samples needed for distinguishing is off by roughly an order of magnitude.
	\item We introduce a pruned FFT method. By only testing probable hypotheses, we improve the performance of the FFT method from~\cite{EC:DucTraVau15} with no computational overhead.
	\item We indicate that the sample dependency due to using LF2 or sample amplification is limited.
	
\end{enumerate}


\subsection{Organization}
\label{sec:Organization}
The rest of the paper is organized as follows. Section~\ref{sec:Background} introduces some necessary background. In Section~\ref{sec:BKW} we cover the basic BKW algorithm. Section~\ref{sec:Distinguishers} goes over distinguishers used for hypothesis testing when solving LWE using BKW and introduces the pruned FFT method. Next, in Section~\ref{sec:FFTequalsOptimal} we show why the FFT distinguisher and the optimal distinguisher perform identically for our setting, followed by simulation results in Section~\ref{sec:Simulation}. Section~\ref{sec:Conclusion} concludes the paper.


\section{Background}
\label{sec:Background}
Let us introduce some notation. Bold small letters denote vectors. Let \(\scalar{\cdot}{\cdot}\) denote the scalar products of two vectors with the same dimension. By \(|x|\) we denote the absolute value of \(x\)  for a real number \(x \in \mathbb{R}\). We also denote by \(\Re{(y)}\) the real part and \(\norm{y}\) the absolute value of a complex number  \(y \in \mathbb{C}\).


\subsection{LWE}
Let us define the LWE problem.

\begin{defn}[LWE]
	
	Let $n$ be a positive integer, $q$ an odd prime. Let $\bf s$ be a uniformly random secret vector in $\Zqn$. Assume access to $m$ noisy scalar products between $\bf s$ and known vectors $\bf a_i$, i.e.
	\begin{equation} \label{eq:LWE}
		b_i = \scalar{{\bf a_i}}{{\bf s}} + {e_i},
	\end{equation}
	for $i = 1, \ldots, m$. The error terms ${e_i}$ are drawn from a distribution $\chi$. The (search) LWE problem is to find $\bf s$.
	
\end{defn}

Thus, when solving LWE you have access to a large set of pairs $(\mathbf{a_i}, b_i)$ and want to find the corresponding secret vector $\mathbf{s}$. Some versions restrict the number of available samples. If we let $\mathbf{b} = (b_1, b_2, \ldots, b_m)$, $\mathbf{e} = (e_1, e_2, \ldots, e_m)$ and $\mathbf{A} = [ \mathbf{a}_1^T, \mathbf{a}_2^T \cdots \mathbf{a}_m^T  ]$ we can write the problem on matrix form as

\begin{equation} \label{eq:LWEMatrix}
\mathbf{b} = \mathbf{s} \mathbf{A} + \mathbf{e}.
\end{equation}

\subsection{Rounded Gaussian Distribution}
For the error we use the rounded Gaussian distribution\footnote{Also common is to use the Discrete Gaussian distribution, which is similar.}. Let $f(x|0, \sigma^2)$ denote the PDF of the normal ditribution with mean 0 and standard deviation $\sigma$, this distribution in turn being denoted as $\mathcal{N}(0, \sigma^2)$. The rounded Gaussian distribution samples from $\mathcal{N}(0, \sigma^2)$, rounds to the nearest integer and wraps to the interval $[-(q - 1)/2, (q - 1)/2]$. In other words, the probability of choosing a certain error $e$ is equal to

\begin{equation*}
\sum_{k = -\infty}^{\infty} \int_{e - 1/2 + k \cdot q}^{e + 1/2 + k \cdot q} f(x|0, \sigma^2) dx,
\end{equation*}

for $e \in [-(q - 1)/2, (q - 1)/2]$. We denote this distribution by $\bar{\Psi}_{\sigma, q}$. We use the well-known heuristic approximation that the sum of two independent distributions $X_1$ and $X_2$, drawn from $\bar{\Psi}_{\sigma_1, q}$ and $\bar{\Psi}_{\sigma_2, q}$, is drawn from $\bar{\Psi}_{\sqrt{\sigma_1^2 + \sigma_2^2}, q}$. We also use the notation $\alpha = \sigma/q$. Finally, we let $\mathcal{U}(a, b)$ denote the discrete uniform distribution taking values from $a$ up to $b$.


\section{BKW}
\label{sec:BKW}
The BKW algorithm was originally invented to solve LPN. It was first used for LWE in \cite{Albrecht2015}.
The BKW algorithm consists of two parts, reduction and hypothesis testing.

\subsection{Reduction}
We divide samples into categories based on $b$ position values in the $\mathbf{a}$ vectors. Two samples should be in the same category if and only if the $b$ position values get canceled when adding or subtracting the $\mathbf{a}$ vectors. Given two samples $([\pm \mathbf{a_0}, \mathbf{a_1}], b_1)$ and $([\pm \mathbf{a_0}, \mathbf{a_2}], b_2)$ within the same category. By adding/subtracting the $\mathbf{a}$ vectors we get

$$\mathbf{a_{1, 2}} = [\underbrace{\begin{matrix} 
	0&0&\cdots&0
	\end{matrix}}_{b \textnormal{ symbols}}\begin{matrix}
&*&*&\cdots&*
\end{matrix}].$$

The corresponding $b$ value is $b_{1, 2} = b_1 \pm b_2$. Now we have a new sample $(\mathbf{a_{1, 2}}, b_{1, 2})$. The corresponding noise variable is $e_{1, 2} = e_1 \pm e_2$, with variance $2\sigma^2$, where $\sigma^2$ is the variance of the originial noise. By calculating a suitable number of new samples for each category we have reduced the dimensionality of the problem by $b$, but increased the noise variance to $2\sigma^2$. If we repeat the reduction process $t$ times we end up with a dimensionality of $n - tb$, and a noise variance of $2^{t} \cdot \sigma^2$.

\subsubsection{LF1 and LF2}
LF1 and LF2 are two implementation tricks originally proposed for solving LPN in~\cite{SCN:LevFou06}. Both can naturally be generalized for solving LWE.

In LF1 we choose one representative per category. We form new samples by the other samples with the representative. This way all samples at the hypothesis testing stage are independent of each other. However, the sample size shrinks by $(q^b - 1)/2$ samples per generation, requiring a large initial sample size.

In LF2 we allow combining any pair of samples within a category, creating much more samples. If we form every possible sample, a sample size of $3(q^b - 1)/2$ is enough to keep the sample size constant between steps. The disadvantage of this approach is that the samples are no longer independent, leading to higher noise levels in the hypothesis stage of BKW. It is generally assumed that this effect is quite small. This assumption is well tested for solving the LPN problem~\cite{SCN:LevFou06}.

\subsubsection{Sample Amplification}
Some versions of LWE limit the number of samples. We can get more samples using sample amplification. For example, by adding/subtracting triples of samples we can increase the initial sample size $m$ up to a maximum of $4 \cdot \binom{m}{3}$. This does increase the noise by a factor of $\sqrt{3}$. It also leads to an increased dependency between samples in the hypothesis testing phase, similar in principle to LF2.


\subsubsection{Secret-Noise Transformation}
\label{sec:Transformation}
There is a transformation of the LWE problem that makes the distribution of the secret vector follow the distribution of the noise~{\cite{EPRINT:Kirchner11, C:ACPS09}. 

\subsubsection{Improved Reduction Steps}
There are many improvements of the plain BKW steps. Lazy modulus switching (LMS) was introduced in \cite{PKC:AFFP14} and further developed in~\cite{C:KirFou15}. In \cite{C:GuoJohSta15} coded-BKW was introduced. Coded-BKW with sieving was introduced in \cite{AC:GJMS17} and improved in \cite{DBLP:journals/tit/GuoJMW19, DBLP:conf/isit/Martensson19}.

Since only the final noise level, not the type of steps, matters for the distinguishers, we only use plain steps in this paper.


\subsection{Hypothesis Testing}
Assume that we have reduced all but $k$ positions to 0, leaving $k$ positions for the hypothesis testing phase. After the reduction phase we have samples on the form

\begin{equation} \label{eq:LWEDistinguish}
b = \sum_{i=1}^{k} a_i \cdot s_i + e \Leftrightarrow b - \sum_{i=1}^{k} a_i \cdot s_i = e,
\end{equation}

where $e$ is (approximately) rounded Gaussian distributed with a standard deviation of $\sigma_f = 2^{t/2} \cdot \sigma$ and mean 0. Now the problem is to distinguish the correct guess $\mathbf{s} = (s_1, s_2, \ldots, s_k)$ from all the incorrect ones, among all $q^k$ guesses\footnote{After the secret-noise transforming most of these hypotheses are almost guaranteed to be incorrect, simplifying the hypothesis testing a bit.}. For each guess $\hat{\mathbf{s}}$ we calculate the corresponding error terms in~\eqref{eq:LWEDistinguish}. For the correct guess the observed values of $e$ are rounded Gaussian distributed, while for the wrong guess they are uniformly random. How to distinguish the right guess from all the wrong ones is explained in Section~\ref{sec:Distinguishers}.


\section{Distinguishers}
\label{sec:Distinguishers}

For the hypothesis testing we study the optimal distinguisher, which is an exhaustive search method; and a faster method based on the fast Fourier transform.


\subsection{Optimal Distinguisher}
\label{sec:OptDist}
Let $D_{\hat{\mathbf{s}}}$ denote the distribution of the $e$ values for a given guess of the secret vector $\hat{\mathbf{s}}$. As is shown in~\cite[Prop. 1]{AC:BaiJunVau04} to optimally distinguish the hypothesis $D_{\hat{\mathbf{s}}} = \mathcal{U}(0, q - 1)$ against $D_{\hat{\mathbf{s}}} = \bar{\Psi}_{\sigma_f, q}$ we calculate the log-likelihood ratio

\begin{equation} \label{eq:LLR}
	\sum_{e=0}^{q - 1} N(e) \log \frac{\Pr_{\bar{\Psi}_{\sigma_f, q}}(e)}{\Pr_{\mathcal{U}(0, q - 1)}(e)} = \sum_{e=0}^{q - 1} N(e) \log \left( q \cdot \Pr\nolimits_{\bar{\Psi}_{\sigma_f, q}}(e) \right),
\end{equation}

where $N(e)$ denotes the number of times $e$ occurs for the guess $\hat{\mathbf{s}}$, $\sigma_f$ denotes the standard deviation of the samples after the reduction phase and $\Pr_D(e)$ denotes the probability of drawing $e$ from the distribution $D$. We choose the value $\hat{\mathbf{s}}$ that maximizes \eqref{eq:LLR}. The time complexity of this distinguisher is

\begin{equation} \label{eq:Exhaustive}
	\mathcal{O}(m \cdot q^k),
\end{equation}
if we try all possible hypotheses. After performing the secret-noise transformation of Section~\ref{sec:Transformation} we can limit ourselves to assuming that the $k$ values in $\mathbf{s}$ have an absolute value of at most $d$, reducing the complexity to

\begin{equation} \label{eq:ExhaustiveLimited}
	\mathcal{O}(m \cdot (2d + 1)^k).
\end{equation}

By only testing the likely hypotheses we have a lower risk of choosing an incorrect one\footnote{as long as the correct one is among our hypotheses.}. This trick of limiting the number of hypotheses can of course also be applied to the FFT method of Section~\ref{sec:FFT}, which we do in Section~\ref{sec:PrunedFFT}.

%
%



\subsection{Fast Fourier Transform Method}
\label{sec:FFT}
For LWE, the idea of using a transform to speed up the distinguishing was introduced in \cite{EC:DucTraVau15}. Consider the function

\begin{equation} \label{eq:f}
	f(\mathbf{x}) = \sum_{j = 1}^{m} \mathds{1}_{\mathbf{a}_j = \mathbf{x}} \theta_q^{b_j},
\end{equation}
where $\mathbf{x} \in \mathbb{Z}_q^k$, $\mathds{1}_{\mathbf{a}_j = \mathbf{x}}$ is equal to 1 if and only if $\mathbf{x} = \mathbf{a}_j$ and 0 otherwise, and $\theta_q$ denotes the $q$-th root of unity. The idea of the FFT distinguisher is to calculate the FFT of $f$, that is

\begin{equation} \label{eq:FFT}
	\hat{f}(\boldsymbol{\alpha}) = \sum_{\mathbf{x} \in \mathbb{Z}_q^k} f(\mathbf{x}) \theta_q^{-\scalar{\mathbf{x}}{\boldsymbol{\alpha}}} = \sum_{j = 1}^{m} \theta_q^{-(\scalar{\mathbf{a}_j}{\boldsymbol{\alpha}} - b_j)}.
\end{equation}

Given enough samples compared to the noise level, the correct guess $\boldsymbol{\alpha} = \mathbf{s}$ maximizes $\Re(\hat{f}(\boldsymbol{\alpha}))$ in \eqref{eq:FFT}.

The time complexity of the FFT distinguisher is

\begin{equation} \label{eq:FFTComplexity} 
\mathcal{O}(m + k \cdot q^k \cdot \log(q)).  
\end{equation}

In general this complexity is much lower than the one in \eqref{eq:Exhaustive}. However, it does depend on the sparsity of the secret $\mathbf{s}$. For a binary $\mathbf{s}$, the exhaustive methods are better.

From~\cite[Thm. 16]{EC:DucTraVau15} we have the following (upper limit) formula for the sample complexity of the FFT distinguisher 

\begin{equation} \label{eq:SamplesTheory}
8 \cdot \ln\left(\frac{q^k}{\epsilon}\right) \left( \frac{q}{\pi} \sin\left(\frac{\pi}{q}\right)  e^{-2\pi^2\sigma^2/q^2}\right)^{-2^{t + 1}},
\end{equation}

where $\epsilon$ is the probability of guessing $\mathbf{s}$ incorrectly. Notice that the expression is slightly modified to fit our notation and that a minor error in the formula is corrected\footnote{Using our notation $k$ should be within the logarithm and not as a factor in front of it like in \cite{EC:DucTraVau15}.}. 

 \subsection{Polynomial Reconstruction Method}
\label{subsec:hybrid}

In~\cite{C:GuoJohSta15}, a method combining exhaustive search and the FFT was introduced.
It achieves optimal distinguishing information theoretically, while being more efficient than the optimal distinguisher. However, its complexity is roughly a factor $q$ higher than the complexity of the FFT distinguisher.


\subsection{Pruned FFT Distinguisher}
\label{sec:PrunedFFT}
Also when using an FFT distinguisher we can limit the number of hypotheses. We only need a small subset of the output values of the FFT distinguisher in \eqref{eq:FFT}, so we can speed-up the calculations using a pruned FFT. In general, if we only need $K$ out of all $N$ output values, the time complexity for calculating the FFT improves from $\mathcal{O}(N \log(N))$ to $\mathcal{O}(N \log(K))$~\cite{PrunedFFT}. Limiting the magnitude when guessing the last $k$ positions of $\mathbf{s}$ to $d$, this changes the time complexity from \eqref{eq:FFTComplexity} to

 \begin{equation} \label{eq:FFTComplexityPruned} 
 \mathcal{O}(m + k \cdot q^k \cdot \log(2d + 1)).  
 \end{equation}

More importantly this method reduces the sample complexity. In the formula for sample complexity \eqref{eq:SamplesTheory}, the numerator $q^k$ corresponds to the number of values of $\mathbf{s}$ can take on the last $k$ positions. Re-doing the proofs of~\cite[Thm. 16]{EC:DucTraVau15}, limiting the magnitude of the guess in each position to $d$, we get

\begin{equation} \label{eq:SamplesLimHypTheory}
8 \cdot \ln\left(\frac{(2d + 1)^k}{\epsilon}\right) \left( \frac{q}{\pi} \sin\left(\frac{\pi}{q}\right)  e^{-2\pi^2\sigma^2/q^2}\right)^{-2^{t + 1}}.
\end{equation}

This reduced sample complexity comes at no extra cost.

\section{Equal Performance of Optimal and FFT Distinguishers}
\label{sec:FFTequalsOptimal}
 
\iftoggle{full}{%
	When starting to run simulations, we noticed that the FFT distinguisher and the optimal distinguisher performed identically, in terms of number of samples to correctly guess the secret. We explain this phenomenon in Appendix~\ref{sec:FFTisOptimal}\footnote{We do, of course, not claim that this is true in general for distinguishing distributions outside of our context of solving LWE using BKW.}.
}{
	When starting to run simulations, we noticed that the FFT distinguisher and the optimal distinguisher performed identically, in terms of number of samples to correctly guess the secret. We explain this phenomenon in Appendix A of~\cite{gms2021}\footnote{We do, of course, not claim that this is true in general for distinguishing distributions outside of our context of solving LWE using BKW.}.
}

There are two immediate effects of this finding.

\begin{itemize}
	\item The polynomial reconstruction method is obsolete.
	\item Unless the secret is very sparse, the FFT distinguisher is strictly better than the optimal distinguisher, since it is computationally cheaper.
\end{itemize}

Hence we limit our investigation to the FFT distinguisher from Section~\ref{sec:Simulation}. We do not make any claims about the equivalance between the sample complexity of the two distinguishers outside of our context of solving LWE using BKW, when having large rounded (or Discrete) Gaussian noise \footnote{Although it could be interesting to investigate.}.




\section{Simulations and Results}
\label{sec:Simulation}
\captionsetup[figure]{font=small}

\iftoggle{full}{%
	This section covers the simulations we ran, using the FBBL library~\cite{FBBL} from~\cite{IndoBKW}, and the results they yielded. For all figures, each point corresponds to running plain BKW plus distinguishing at least 30 times. For most points we ran slightly more iterations. See Appendix~\ref{sec:SimulationDetails} for details on the number of iterations for all the points. We chose our parameters inspired by the Darmstadt LWE Challenge~\cite{DarmstadtLWE}.
}{
	This section covers the simulations we ran, using the FBBL library~\cite{FBBL} from~\cite{IndoBKW}, and the results they yielded. For all figures, each point corresponds to running plain BKW plus distinguishing at least 30 times. For most points we ran slightly more iterations. See Appendix~B of~\cite{gms2021} for details on the number of iterations for all the points. We chose our parameters inspired by the Darmstadt LWE Challenge~\cite{DarmstadtLWE}.
}

The challenges are a set of (search) LWE instances used to compare LWE solving methods. Each instance consists of the dimension $n$, the modulus $q \approx n^2$, the relative error size $\alpha$ and $m \approx n^2$ equations of the form~\eqref{eq:LWE}. Our simulations mostly use parameters inspired by the LWE challenges. We mostly let $q = 1601$ (corresponding to $n = 40$) and vary $\alpha$ to get problem instances that require a suitable number of samples for simulating hypothesis testing. The records for the LWE challenges are set using lattice sieving~\cite{EC:ADHKPS19}.

\subsection{Varying Noise Level}
\label{sec:SimComparisonAlpha}

In the upper part of Figure \ref{fig:theoryVsSimulation} we compare the theoretical sample complexity from \eqref{eq:SamplesTheory} with simulation results from an implementation of the FFT distinguisher of \cite{EC:DucTraVau15} and our pruned FFT distinguisher. The latter distinguisher guesses values of absolute value up to $3\sigma$, rounded upwards. The simulated points are the median values of our simulations and the theoretical values correspond to setting $\epsilon = 0.5$ in \eqref{eq:SamplesTheory}. We use $q = 1601$, $n = 28$, we take $t = 13$ steps of plain BKW, reducing 2 positions per step. Finally we guess the last 2 positions and measure the minimum number of samples to correctly guess the secret. We vary $\alpha$ between 0.005 and 0.006. We use LF1 to guarantee that the samples are independent. 


We notice that there is a gap of roughly a factor 10 between theory and simulation. More exactly, the gap is a factor [10.8277, 8.6816, 10.1037, 8.6776, 10.5218, 10.1564] for the six points, counting in increasing order of noise level.

We also see a gap between the FFT distinguisher and pruned FFT distinguisher. We can estimate the gap by comparing \eqref{eq:SamplesLimHypTheory} and \eqref{eq:SamplesTheory}. Counting in increasing level of noise by theory we expect the pruned version to need [1.8056, 1.8056, 1.7895, 1.7743, 1.7598, 1.7461] times less samples for the 6 data points. The numbers from the simulation were [2.0244, 1.8610, 1.8433, 2.1905, 2.0665, 2.2060], pretty close to theory.

\subsection{Varying $q$}
\label{sec:SimComparisonQ}

In the lower part of Figure \ref{fig:theoryVsSimulation} we show how the number of samples needed for distinguishing varies with $q$. For $q$ we use the values [101, 201, 401, 801, 1601, 3201], for $\alpha$ we use the values [0.0896, 0.0448, 0.0224, 0.0112, 0.0056, 0.0028] and the number of steps were [5, 7, 9, 11, 13, 15]. Thereby the final noise level and the original $\mathbf{s}$ vectors have almost the same distribution, making the $q$ values the only varying factor. We use LF1 to guarantee that the samples are independent.


Notice that the number of samples needed to guess the secret is roughly an order of magnitude lower than theory predicts, counting in increasing order of $q$, the gain is a factor [11.4537, 10.6112, 9.2315, 10.4473, 9.5561, 9.7822] for the six points.

Also notice that the pruned version is an improvement, that increases with $q$. This is because the total number of hypotheses divided by the number of hypotheses we make increases with $q$. By comparing \eqref{eq:SamplesLimHypTheory} and \eqref{eq:SamplesTheory}, we expect the improvement to be a factor [1.1303, 1.2871, 1.4563, 1.6152, 1.7743, 1.9334]. This is pretty close to the factors 1.1435, 1.4551, 1.6215, 1.8507, 2.0121, 2.3045] from simulation.

\begin{figure}[]
	\begin{center}
		\begin{subfigure}{.36\textwidth}
			\centering
			\includegraphics[width=\linewidth]{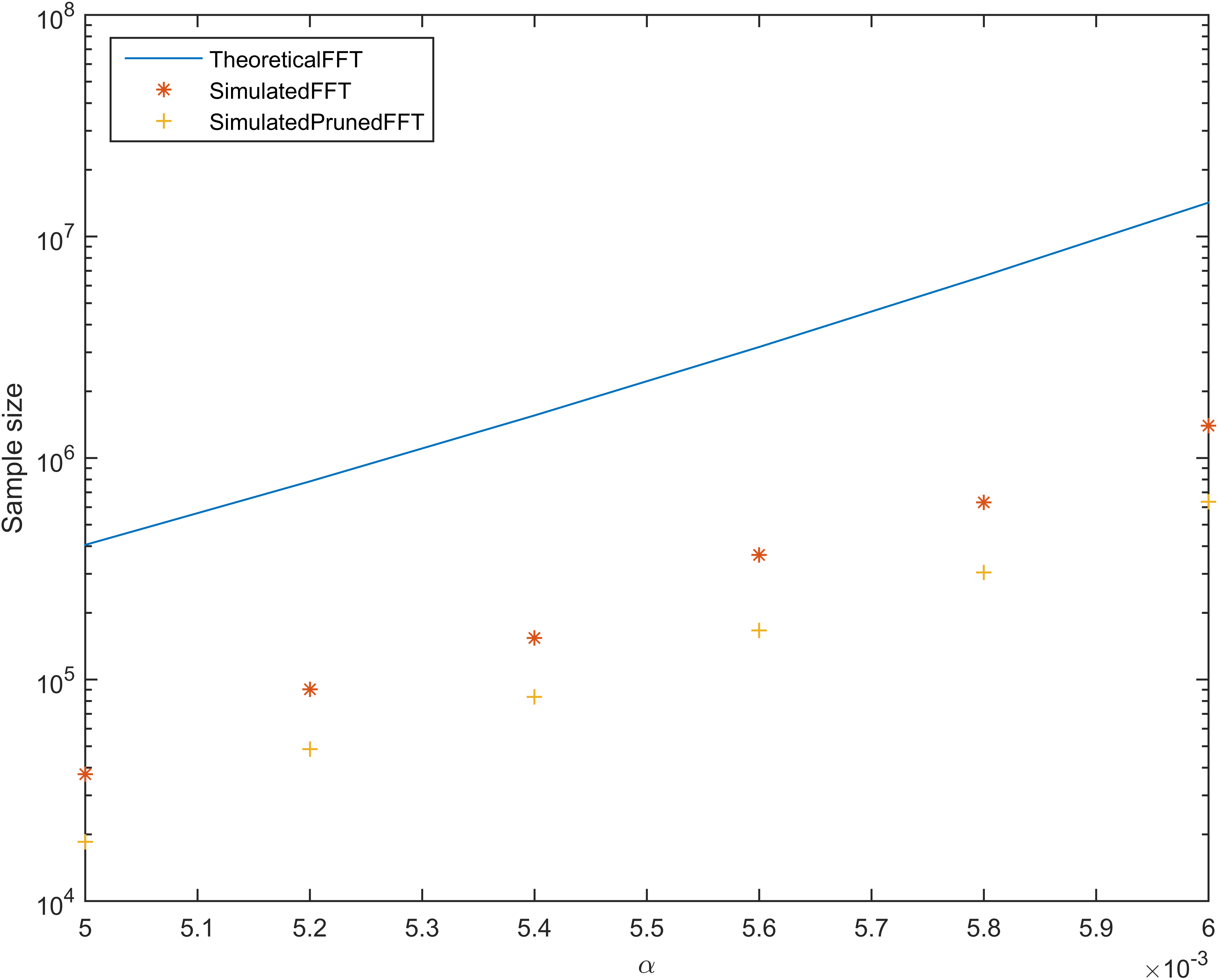}  
			\label{fig:VaryingAlpha}
			\vspace{-11pt}
		\end{subfigure}
		\begin{subfigure}{.35\textwidth}
			\centering
			\includegraphics[width=\linewidth]{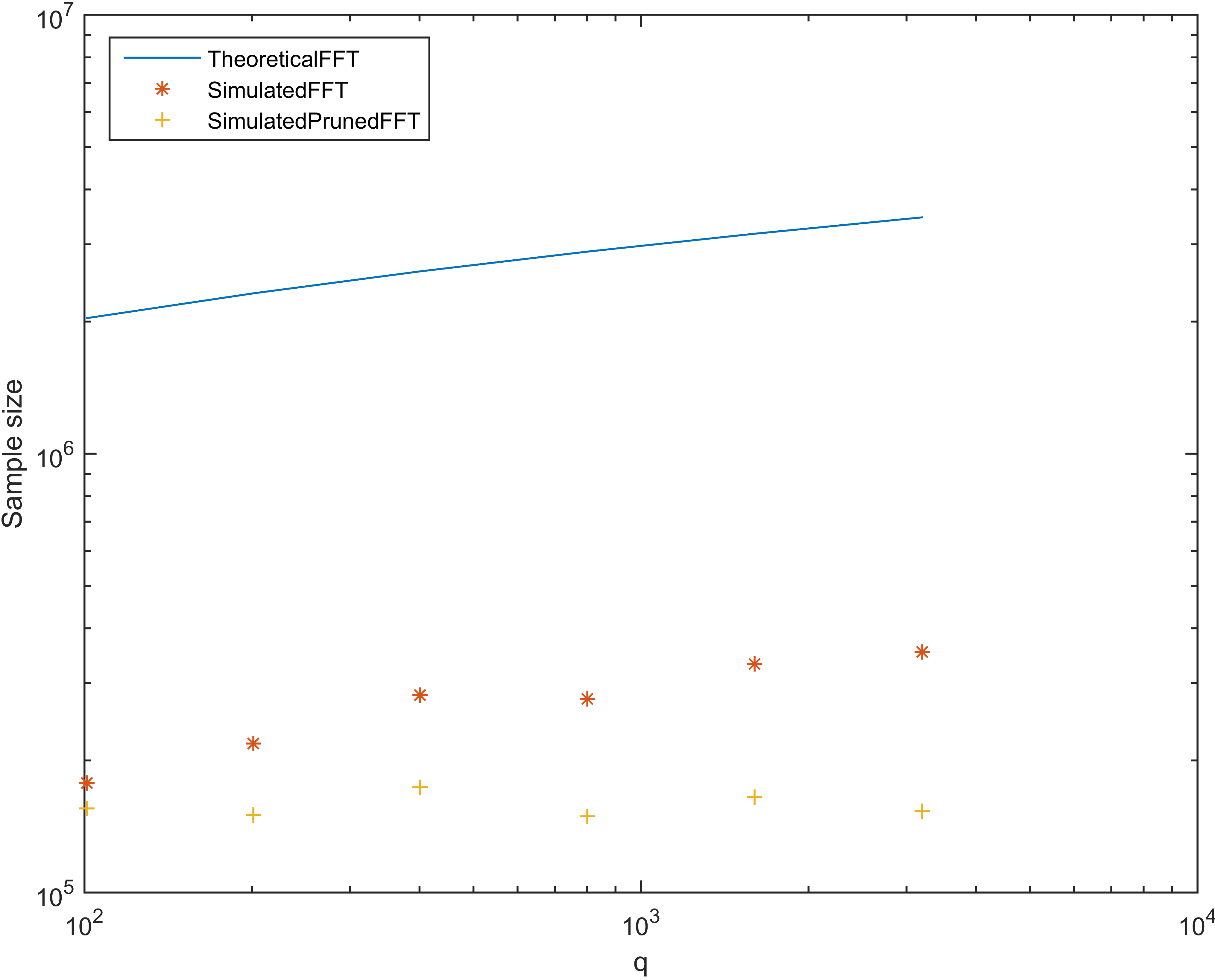}  
			\label{fig:VaryingQ}
			\vspace{-18pt}
		\end{subfigure}
		\caption{Theoretical values vs. simulated values.}
		\label{fig:theoryVsSimulation} 
		\vspace{-12pt}
	\end{center}
\end{figure}

\subsection{LF1 vs LF2}
\label{sec:SimLF1LF2}

We investigate the increased number of samples needed due to dependencies, when using LF2. For LF2, depending on the number of samples needed for guessing, we used either the minimum number of samples to produce a new generation of the same size or a sample size roughly equal to the size needed for guessing at the end. To test the limit of LF2 we made sure to produce every possible sample from each category. See the upper part of Figure~\ref{fig:sampleDependence} for details. The setting is the same as in Section~\ref{sec:SimComparisonAlpha}. We only use the pruned FFT distinguisher. Notice that the performance is almost exactly the same in both the LF1 and the LF2 cases, as is generally assumed~\cite{SCN:LevFou06}.

%
%

\subsection{Sample Amplification}
\label{sec:SimAmplification}

The lower part of Figure~\ref{fig:sampleDependence} shows the increased number of samples needed, due to sample amplification. We use $q = 1601$ and 1600 initial samples. We form new samples by combining triples of samples to get a large enough sample size. We vary the noise level between $\alpha = 0.005/\sqrt{3}$ and $\alpha = 0.006/\sqrt{3}$. We take 13 steps of plain BKW, reducing 2 positions per step. Finally we guess the last 2 positions and measure the minimum number of samples needed to guess correctly. We use LF1 and we compare the results against starting with as many samples as we want and noise levels between $\alpha = 0.005$ and $\alpha = 0.006$, both tricks to isolate the dependency due to sample amplification. We only use the pruned FFT distinguisher. The difference between the points is small, implying that the dependency due to sample amplification is limited.

%
%


\begin{figure}[h!]
	\begin{center}
		\begin{subfigure}{.36\textwidth}
			\centering
			\includegraphics[width=\linewidth]{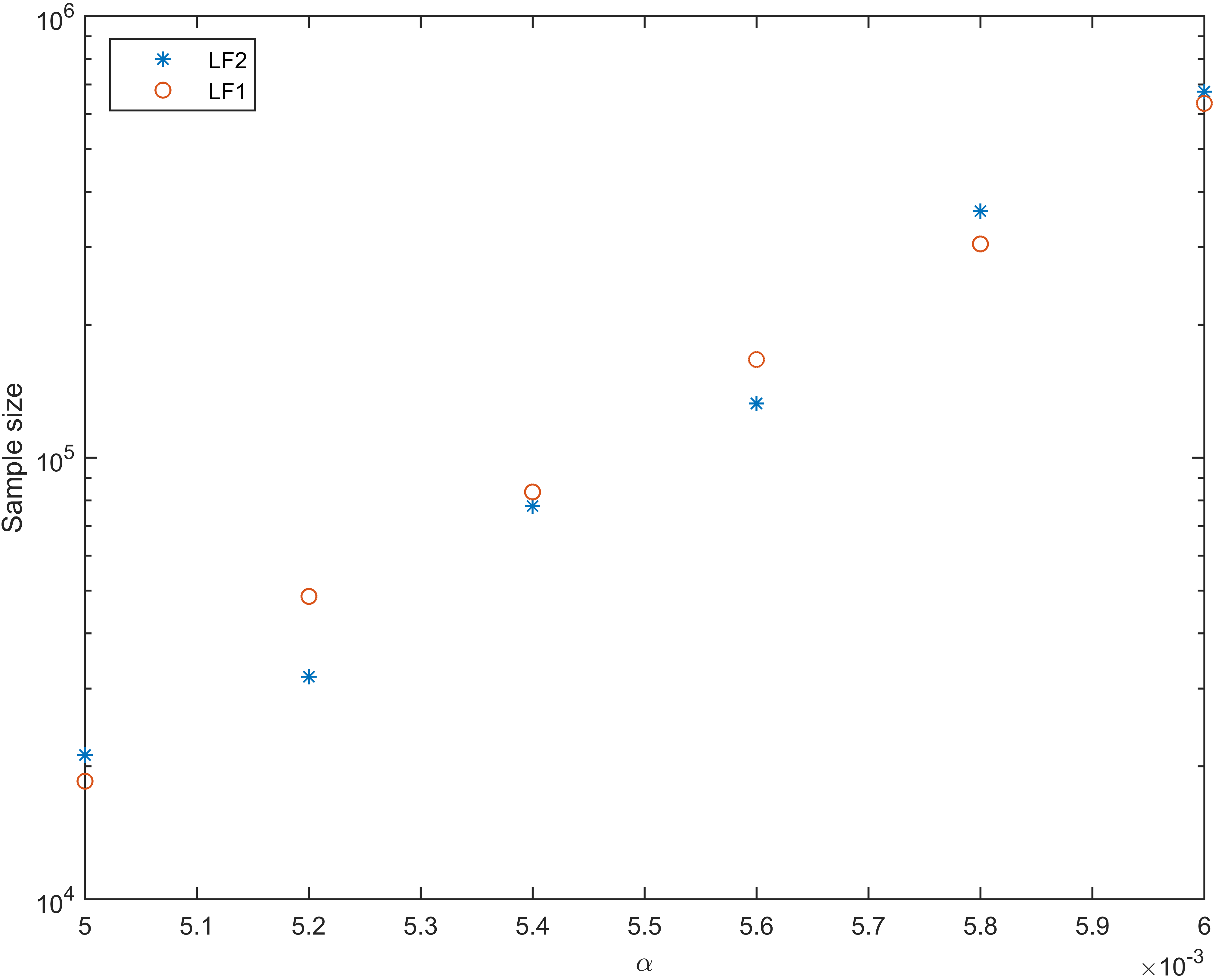}  
			\label{fig:LF1LF2}
			\vspace{-11pt}
		\end{subfigure}
		\begin{subfigure}{.35\textwidth}
			\centering
			\includegraphics[width=\linewidth]{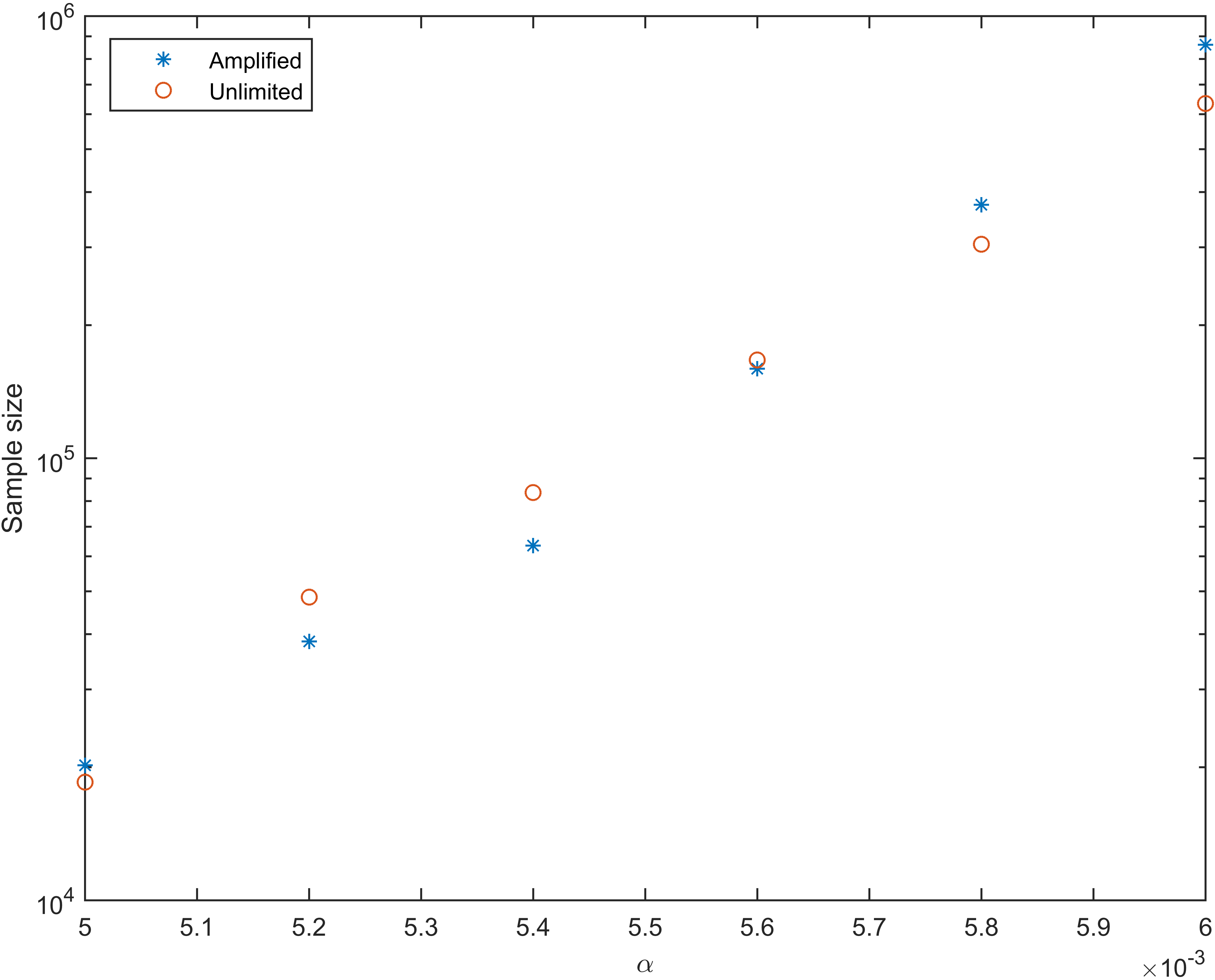}  
			\label{fig:Amplification}
			\vspace{-18pt}
		\end{subfigure}
		\caption{Testing the effect of sample dependence.}
		\label{fig:sampleDependence}
		\vspace{-16pt}
	\end{center}
\end{figure}


 







 
\section{Conclusions}
\label{sec:Conclusion}

We have shown that the FFT distinguisher and the optimal distinguisher have the same sample complexity for solving LWE using BKW. We have also showed that it performs roughly an order of magnitude better than the upper limit formula from \cite[Thm. 16]{EC:DucTraVau15}. Our pruned version of the FFT method improves the sample complexity of the FFT solver, at no cost. Finally, we have indicated that the sample dependency due to both LF2 and sample amplification is limited.



\bibliographystyle{IEEEtran}
\bibliography{abbrev0,crypto,other}

\begin{thebibliography}{10}
\providecommand{\url}[1]{#1}
\csname url@samestyle\endcsname
\providecommand{\newblock}{\relax}
\providecommand{\bibinfo}[2]{#2}
\providecommand{\BIBentrySTDinterwordspacing}{\spaceskip=0pt\relax}
\providecommand{\BIBentryALTinterwordstretchfactor}{4}
\providecommand{\BIBentryALTinterwordspacing}{\spaceskip=\fontdimen2\font plus
\BIBentryALTinterwordstretchfactor\fontdimen3\font minus
  \fontdimen4\font\relax}
\providecommand{\BIBforeignlanguage}[2]{{%
\expandafter\ifx\csname l@#1\endcsname\relax
\typeout{** WARNING: IEEEtran.bst: No hyphenation pattern has been}%
\typeout{** loaded for the language `#1'. Using the pattern for}%
\typeout{** the default language instead.}%
\else
\language=\csname l@#1\endcsname
\fi
#2}}
\providecommand{\BIBdecl}{\relax}
\BIBdecl

\bibitem{FOCS:Shor94}
P.~W. Shor, ``Algorithms for quantum computation: Discrete logarithms and
  factoring,'' in \emph{35th Annual Symposium on Foundations of Computer
  Science}.\hskip 1em plus 0.5em minus 0.4em\relax Santa Fe, NM, USA: {IEEE}
  Computer Society Press, Nov.~20--22, 1994, pp. 124--134.

\bibitem{NISTPQCweb}
``{NIST Post-Quantum Cryptography Standardization},''
  \url{https://csrc.nist.gov/Projects/Post-Quantum-Cryptography/Post-Quantum-Cryptography-Standardization},
  accessed: 2019-09-24.

\bibitem{STOC:Regev05}
O.~Regev, ``On lattices, learning with errors, random linear codes, and
  cryptography,'' in \emph{37th Annual {ACM} Symposium on Theory of Computing},
  H.~N. Gabow and R.~Fagin, Eds.\hskip 1em plus 0.5em minus 0.4em\relax
  Baltimore, MA, USA: {ACM} Press, May~22--24, 2005, pp. 84--93.

\bibitem{C:BFKL93}
A.~Blum, M.~L. Furst, M.~J. Kearns, and R.~J. Lipton, ``Cryptographic
  primitives based on hard learning problems,'' in \emph{Advances in Cryptology
  -- {CRYPTO}'93}, ser. Lecture Notes in Computer Science, D.~R. Stinson, Ed.,
  vol. 773.\hskip 1em plus 0.5em minus 0.4em\relax Santa Barbara, CA, USA:
  Springer, Heidelberg, Germany, Aug.~22--26, 1994, pp. 278--291.

\bibitem{STOC:BluKalWas00}
A.~Blum, A.~Kalai, and H.~Wasserman, ``Noise-tolerant learning, the parity
  problem, and the statistical query model,'' in \emph{32nd Annual {ACM}
  Symposium on Theory of Computing}.\hskip 1em plus 0.5em minus 0.4em\relax
  Portland, OR, USA: {ACM} Press, May~21--23, 2000, pp. 435--440.

\bibitem{DBLP:journals/jacm/BlumKW03}
\BIBentryALTinterwordspacing
------, ``Noise-tolerant learning, the parity problem, and the statistical
  query model,'' \emph{J. {ACM}}, vol.~50, no.~4, pp. 506--519, 2003. [Online].
  Available: \url{https://doi.org/10.1145/792538.792543}
\BIBentrySTDinterwordspacing

\bibitem{ConcreteLWE}
M.~R. Albrecht, R.~Player, and S.~Scott, ``{On The Concrete Hardness Of
  Learning With Errors},'' \emph{J. Mathematical Cryptology}, vol.~9, no.~3,
  pp. 169--203, 2015.

\bibitem{DBLP:journals/dcc/HeroldKM18}
\BIBentryALTinterwordspacing
G.~Herold, E.~Kirshanova, and A.~May, ``On the asymptotic complexity of solving
  {LWE},'' \emph{Des. Codes Cryptogr.}, vol.~86, no.~1, pp. 55--83, 2018.
  [Online]. Available: \url{https://doi.org/10.1007/s10623-016-0326-0}
\BIBentrySTDinterwordspacing

\bibitem{DBLP:journals/tit/GuoJMW19}
\BIBentryALTinterwordspacing
Q.~Guo, T.~Johansson, E.~M{\aa}rtensson, and P.~{Stankovski Wagner}, ``On the
  asymptotics of solving the {LWE} problem using coded-bkw with sieving,''
  \emph{{IEEE} Trans. Information Theory}, vol.~65, no.~8, pp. 5243--5259,
  2019. [Online]. Available: \url{https://doi.org/10.1109/TIT.2019.2906233}
\BIBentrySTDinterwordspacing

\bibitem{EC:DucTraVau15}
A.~Duc, F.~Tram{\`e}r, and S.~Vaudenay, ``Better algorithms for {LWE} and
  {LWR},'' in \emph{Advances in Cryptology -- {EUROCRYPT}~2015, Part~I}, ser.
  Lecture Notes in Computer Science, E.~Oswald and M.~Fischlin, Eds., vol.
  9056.\hskip 1em plus 0.5em minus 0.4em\relax Sofia, Bulgaria: Springer,
  Heidelberg, Germany, Apr.~26--30, 2015, pp. 173--202.

\bibitem{SCN:LevFou06}
{\'E}.~Levieil and P.-A. Fouque, ``An improved {LPN} algorithm,'' in \emph{SCN
  06: 5th International Conference on Security in Communication Networks}, ser.
  Lecture Notes in Computer Science, R.~D. Prisco and M.~Yung, Eds., vol.
  4116.\hskip 1em plus 0.5em minus 0.4em\relax Maiori, Italy: Springer,
  Heidelberg, Germany, Sep.~6--8, 2006, pp. 348--359.

\bibitem{EPRINT:Kirchner11}
P.~Kirchner, ``Improved generalized birthday attack,'' Cryptology ePrint
  Archive, Report 2011/377, 2011, \url{http://eprint.iacr.org/2011/377}.

\bibitem{C:ACPS09}
B.~Applebaum, D.~Cash, C.~Peikert, and A.~Sahai, ``Fast cryptographic
  primitives and circular-secure encryption based on hard learning problems,''
  in \emph{Advances in Cryptology -- {CRYPTO}~2009}, ser. Lecture Notes in
  Computer Science, S.~Halevi, Ed., vol. 5677.\hskip 1em plus 0.5em minus
  0.4em\relax Santa Barbara, CA, USA: Springer, Heidelberg, Germany,
  Aug.~16--20, 2009, pp. 595--618.

\bibitem{EPRINT:BerLan12d}
D.~J. Bernstein and T.~Lange, ``Never trust a bunny,'' Cryptology ePrint
  Archive, Report 2012/355, 2012, \url{http://eprint.iacr.org/2012/355}.

\bibitem{AC:GuoJohLon14}
Q.~Guo, T.~Johansson, and C.~L{\"o}ndahl, ``Solving {LPN} using covering
  codes,'' in \emph{Advances in Cryptology -- {ASIACRYPT}~2014, Part~I}, ser.
  Lecture Notes in Computer Science, P.~Sarkar and T.~Iwata, Eds., vol.
  8873.\hskip 1em plus 0.5em minus 0.4em\relax Kaoshiung, Taiwan, R.O.C.:
  Springer, Heidelberg, Germany, Dec.~7--11, 2014, pp. 1--20.

\bibitem{DBLP:journals/joc/GuoJL20}
\BIBentryALTinterwordspacing
Q.~Guo, T.~Johansson, and C.~L{\"{o}}ndahl, ``Solving {LPN} using covering
  codes,'' \emph{J. Cryptology}, vol.~33, no.~1, pp. 1--33, 2020. [Online].
  Available: \url{https://doi.org/10.1007/s00145-019-09338-8}
\BIBentrySTDinterwordspacing

\bibitem{EC:ZhaJiaWan16}
B.~Zhang, L.~Jiao, and M.~Wang, ``Faster algorithms for solving {LPN},'' in
  \emph{Advances in Cryptology -- {EUROCRYPT}~2016, Part~I}, ser. Lecture Notes
  in Computer Science, M.~Fischlin and J.-S. Coron, Eds., vol. 9665.\hskip 1em
  plus 0.5em minus 0.4em\relax Vienna, Austria: Springer, Heidelberg, Germany,
  May~8--12, 2016, pp. 168--195.

\bibitem{AC:BogVau16}
S.~Bogos and S.~Vaudenay, ``Optimization of {LPN} solving algorithms,'' in
  \emph{Advances in Cryptology -- {ASIACRYPT}~2016, Part~I}, ser. Lecture Notes
  in Computer Science, J.~H. Cheon and T.~Takagi, Eds., vol. 10031.\hskip 1em
  plus 0.5em minus 0.4em\relax Hanoi, Vietnam: Springer, Heidelberg, Germany,
  Dec.~4--8, 2016, pp. 703--728.

\bibitem{DBLP:journals/ccds/BogosTV16}
\BIBentryALTinterwordspacing
S.~Bogos, F.~Tram{\`{e}}r, and S.~Vaudenay, ``On solving {L} {P} {N} using {B}
  {K} {W} and variants - implementation and analysis,'' \emph{Cryptography and
  Communications}, vol.~8, no.~3, pp. 331--369, 2016. [Online]. Available:
  \url{https://doi.org/10.1007/s12095-015-0149-2}
\BIBentrySTDinterwordspacing

\bibitem{Albrecht2015}
M.~R. Albrecht, C.~Cid, J.-C. Faug{\`e}re, R.~Fitzpatrick, and L.~Perret, ``{On
  the complexity of the BKW algorithm on LWE},'' \emph{Designs, Codes and
  Cryptography}, vol.~74, no.~2, pp. 325--354, 2015.

\bibitem{PKC:AFFP14}
M.~R. Albrecht, J.-C. Faug{\`e}re, R.~Fitzpatrick, and L.~Perret, ``Lazy
  modulus switching for the {BKW} algorithm on {LWE},'' in \emph{PKC~2014: 17th
  International Conference on Theory and Practice of Public Key Cryptography},
  ser. Lecture Notes in Computer Science, H.~Krawczyk, Ed., vol. 8383.\hskip
  1em plus 0.5em minus 0.4em\relax Buenos Aires, Argentina: Springer,
  Heidelberg, Germany, Mar.~26--28, 2014, pp. 429--445.

\bibitem{C:GuoJohSta15}
Q.~Guo, T.~Johansson, and P.~Stankovski, ``Coded-{BKW}: Solving {LWE} using
  lattice codes,'' in \emph{Advances in Cryptology -- {CRYPTO}~2015, Part~I},
  ser. Lecture Notes in Computer Science, R.~Gennaro and M.~J.~B. Robshaw,
  Eds., vol. 9215.\hskip 1em plus 0.5em minus 0.4em\relax Santa Barbara, CA,
  USA: Springer, Heidelberg, Germany, Aug.~16--20, 2015, pp. 23--42.

\bibitem{C:KirFou15}
P.~Kirchner and P.-A. Fouque, ``An improved {BKW} algorithm for {LWE} with
  applications to cryptography and lattices,'' in \emph{Advances in Cryptology
  -- {CRYPTO}~2015, Part~I}, ser. Lecture Notes in Computer Science, R.~Gennaro
  and M.~J.~B. Robshaw, Eds., vol. 9215.\hskip 1em plus 0.5em minus 0.4em\relax
  Santa Barbara, CA, USA: Springer, Heidelberg, Germany, Aug.~16--20, 2015, pp.
  43--62.

\bibitem{AC:GJMS17}
Q.~Guo, T.~Johansson, E.~M{\r a}rtensson, and P.~Stankovski, ``Coded-{BKW} with
  sieving,'' in \emph{Advances in Cryptology -- {ASIACRYPT}~2017, Part~I}, ser.
  Lecture Notes in Computer Science, T.~Takagi and T.~Peyrin, Eds., vol.
  10624.\hskip 1em plus 0.5em minus 0.4em\relax Hong Kong, China: Springer,
  Heidelberg, Germany, Dec.~3--7, 2017, pp. 323--346.

\bibitem{C:EssKubMay17}
A.~Esser, R.~K{\"u}bler, and A.~May, ``{LPN} decoded,'' in \emph{Advances in
  Cryptology -- {CRYPTO}~2017, Part~II}, ser. Lecture Notes in Computer
  Science, J.~Katz and H.~Shacham, Eds., vol. 10402.\hskip 1em plus 0.5em minus
  0.4em\relax Santa Barbara, CA, USA: Springer, Heidelberg, Germany,
  Aug.~20--24, 2017, pp. 486--514.

\bibitem{C:EHKMS18}
A.~Esser, F.~Heuer, R.~K{\"u}bler, A.~May, and C.~Sohler, ``Dissection-{BKW},''
  in \emph{Advances in Cryptology -- {CRYPTO}~2018, Part~II}, ser. Lecture
  Notes in Computer Science, H.~Shacham and A.~Boldyreva, Eds., vol.
  10992.\hskip 1em plus 0.5em minus 0.4em\relax Santa Barbara, CA, USA:
  Springer, Heidelberg, Germany, Aug.~19--23, 2018, pp. 638--666.

\bibitem{IMA:DelEssMay19}
C.~Delaplace, A.~Esser, and A.~May, ``Improved low-memory subset sum and {LPN}
  algorithms via multiple collisions,'' in \emph{17th IMA International
  Conference on Cryptography and Coding}, ser. Lecture Notes in Computer
  Science, M.~Albrecht, Ed., vol. 11929.\hskip 1em plus 0.5em minus 0.4em\relax
  Oxford, UK: Springer, Heidelberg, Germany, Dec.~16--18, 2019, pp. 178--199.

\bibitem{DBLP:conf/isit/Martensson19}
\BIBentryALTinterwordspacing
E.~M{\aa}rtensson, ``The asymptotic complexity of coded-bkw with sieving using
  increasing reduction factors,'' in \emph{{IEEE} International Symposium on
  Information Theory, {ISIT} 2019, Paris, France, July 7-12, 2019}.\hskip 1em
  plus 0.5em minus 0.4em\relax {IEEE}, 2019, pp. 2579--2583. [Online].
  Available: \url{https://doi.org/10.1109/ISIT.2019.8849218}
\BIBentrySTDinterwordspacing

\bibitem{AC:BaiJunVau04}
T.~Baign{\`e}res, P.~Junod, and S.~Vaudenay, ``How far can we go beyond linear
  cryptanalysis?'' in \emph{Advances in Cryptology -- {ASIACRYPT}~2004}, ser.
  Lecture Notes in Computer Science, P.~J. Lee, Ed., vol. 3329.\hskip 1em plus
  0.5em minus 0.4em\relax Jeju Island, Korea: Springer, Heidelberg, Germany,
  Dec.~5--9, 2004, pp. 432--450.

\bibitem{PrunedFFT}
H.~V. {Sorensen} and C.~S. {Burrus}, ``Efficient computation of the dft with
  only a subset of input or output points,'' \emph{IEEE Transactions on Signal
  Processing}, vol.~41, no.~3, pp. 1184--1200, 1993.

\bibitem{FBBL}
A.~Budroni, E.~Mårtensson, and P.~{Stankovski Wagner}, ``{FBBL} - file-{B}ased
  {BKW} for {LWE},'' \url{https://github.com/FBBL/fbbl}, 2020.

\bibitem{IndoBKW}
A.~Budroni, Q.~Guo, T.~Johansson, E.~M{\aa}rtensson, and P.~S. Wagner, ``Making
  the bkw algorithm practical for lwe,'' in \emph{Progress in Cryptology --
  INDOCRYPT 2020}, K.~Bhargavan, E.~Oswald, and M.~Prabhakaran, Eds.\hskip 1em
  plus 0.5em minus 0.4em\relax Cham: Springer International Publishing, 2020,
  pp. 417--439.

\bibitem{DarmstadtLWE}
``{TU Darmstadt Learning with Errors Challenge},''
  \url{https://www.latticechallenge.org/lwe_challenge/challenge.php}, accessed:
  2020-09-30.

\bibitem{EC:ADHKPS19}
M.~R. Albrecht, L.~Ducas, G.~Herold, E.~Kirshanova, E.~W. Postlethwaite, and
  M.~Stevens, ``The general sieve kernel and new records in lattice
  reduction,'' in \emph{Advances in Cryptology -- {EUROCRYPT}~2019, Part~II},
  ser. Lecture Notes in Computer Science, Y.~Ishai and V.~Rijmen, Eds., vol.
  11477.\hskip 1em plus 0.5em minus 0.4em\relax Darmstadt, Germany: Springer,
  Heidelberg, Germany, May~19--23, 2019, pp. 717--746.

\end{thebibliography}

\iftoggle{full}{%
	\appendices
	\clearpage

\section{Explaining the Optimimality of the FFT Distinguisher}
\label{sec:FFTisOptimal}

Consider a sample on the form \eqref{eq:LWEDistinguish}. By making a guess $\hat{\mathbf{s}}$ we calculate the corresponding error term $\hat{e}$. The Fourier transform of the FFT distinguisher in \eqref{eq:FFT} can now be written as

\begin{equation}
\label{eq:FFTsum}
\sum_{j = 1}^{m} \theta_q^{\hat{e}_j}.
\end{equation}

The real part \eqref{eq:FFTsum} is equal to

\begin{equation}
\label{eq:FFTsumRealPart}
\sum_{j = 1}^{m} \cos{(2 \pi \hat{e}_j / q)}.
\end{equation}

The FFT distinguisher picks the guess that maximizes \eqref{eq:FFTsumRealPart}. Now, let us rewrite \eqref{eq:LLR} for the optimal distinguisher as

\begin{equation} \label{eq:LLRsum}
\sum_{j=1}^{m} \log \left( q \cdot \Pr\nolimits_{\bar{\Psi}_{\sigma_f, q}}(\hat{e}_j) \right).
\end{equation}

It turns out that with increasing noise level, the terms in \eqref{eq:LLRsum} can be approximated as cosine functions with a period of $q$, as illustrated in Figure~\ref{fig:CosineApproximation}. The terms correspond to $q = 1601$, starting with rounded Gaussian noise with $\alpha = 0.005$, $\sigma = \alpha \cdot q = 8.005$ and taking 12 or 13 steps of plain BKW respectively. Notice that the approximation gets drastically better with increasing noise level\footnote{Also notice that the approximation is not necessarily the best cosine approximation. It is simple the approximation that matches the largest and the smallest value of the curve.}. The 13 step picture corresponds to the setting used in most of the experiments in Section~\ref{sec:Simulation}. For a large-scale problem, the noise level would of course be much larger, resulting in an even better cosine approximation.

\begin{figure}[ht]
	\begin{subfigure}{.5\textwidth}
		\centering
		\includegraphics[width=\linewidth]{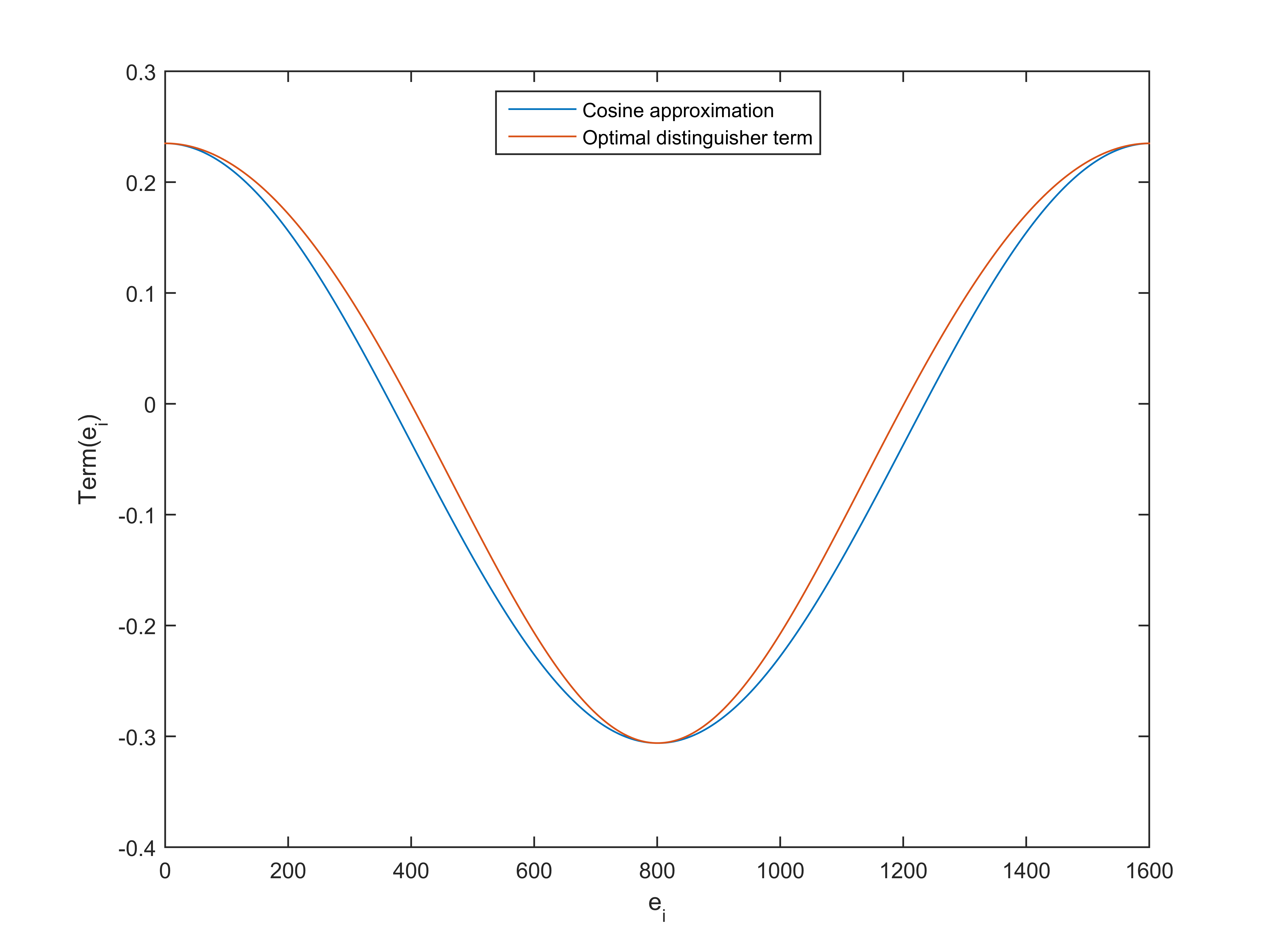}  
		\caption{Taking 12 plain BKW steps}
	\end{subfigure}
	\begin{subfigure}{.5\textwidth}
		\centering
		\includegraphics[width=\linewidth]{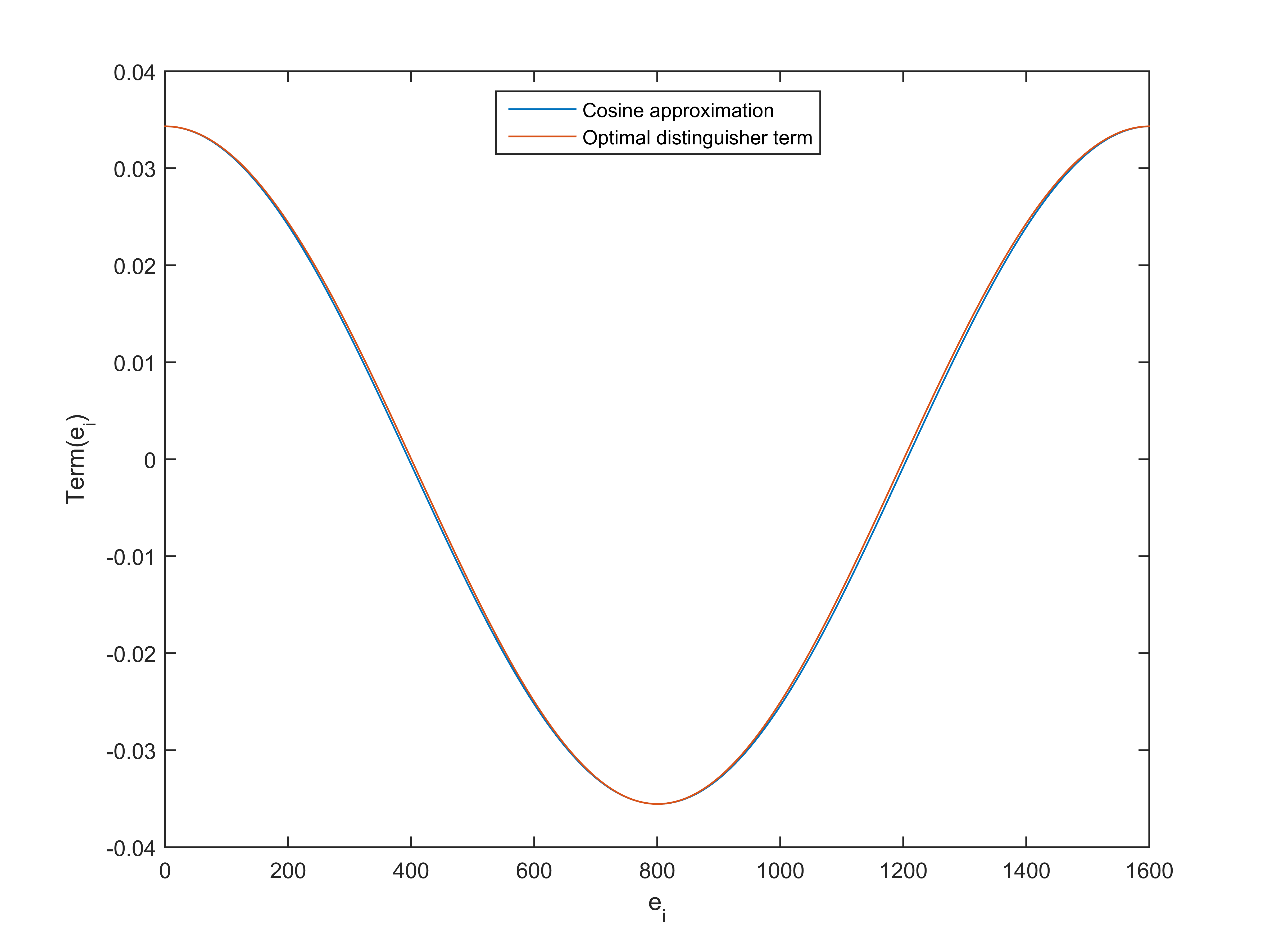}  
		\caption{Taking 13 plain BKW steps}
	\end{subfigure}
	\caption{Approximating the terms in \eqref{eq:LLRsum} as cosine functions.}
	\label{fig:CosineApproximation}
\end{figure}

Since both distinguishers pick the $\hat{\mathbf{s}}$ that minimizes a sum of cosine functions with the same period, they will pick the same $\hat{\mathbf{s}}$, hence they will perform identically.

\clearpage

\section{Number of Iterations in the Simulations}
\label{sec:SimulationDetails}

The following is a collection of lists of the number of iterations used for each point to get the estimations of the median values in Figures~\ref{fig:theoryVsSimulation}-\ref{fig:sampleDependence}. For each figure and curve we list the number iterations from left to right in, in other words in increasing level of noise level $\alpha$ or modulus $q$.

\subsection*{Figure \ref{fig:theoryVsSimulation} - Varying $\alpha$}

\begin{tabular}{ |c|cccccc| } 
	\hline
	Simulated FFT & 31 & 51 & 52 & 59 & 50 & 52 \\
	\hline
	Simulated Pruned FFT & 33 & 41 & 56 & 35 & 30 & 49 \\ 
	\hline
\end{tabular}

\subsection*{Figure \ref{fig:theoryVsSimulation} - Varying $q$}

\begin{tabular}{ |c|cccccc| } 
	\hline
	Simulated FFT & 100 & 100 & 95 & 80 & 67 & 82 \\
	\hline
	Simulated Pruned FFT & 100 & 100 & 95 & 80 & 67 & 82 \\ 
	\hline
\end{tabular}

\subsection*{Figure \ref{fig:sampleDependence} - LF1 vs. LF2}

\begin{tabular}{ |c|cccccc| } 
	\hline
	LF1 & 33 & 41 & 56 & 35 & 30 & 49 \\
	\hline
	LF2 & 43 & 46 & 69 & 37 & 69 & 50 \\ 
	\hline
\end{tabular}

\subsection*{Figure \ref{fig:sampleDependence} - Unlimited vs. Sample Amplification}

\begin{tabular}{ |c|cccccc| } 
	\hline
	Unlimited Samples & 33 & 41 & 56 & 35 & 30 & 49 \\
	\hline
	Sample Amplification & 37 & 59 & 38 & 45 & 47 & 40 \\ 
	\hline
\end{tabular}

}

\end{document}